\documentstyle[aps,version2]{revtex}
\begin{document}
\draft
\begin{title}
K$^+$-nucleus quasielastic scattering
\end{title}
\author{C.~M.~Kormanyos,$^{\rm(1)}$ R.~J.~Peterson,$^{\rm(1)}$
J.~R.~Shepard,$^{\rm(1)}$ J.~E.~Wise,$^{\rm(1)}$
S.~Bart,$^{\rm(2)}$ R.~E.~Chrien,$^{\rm(2)}$ L.~Lee,$^{\rm(3)}$
B.~L.~Clausen,$^{\rm(4)}$ J.~Piekarewicz,$^{\rm(5)}$
M.~B.~Barakat,$^{\rm(6),(a)}$ R.~A.~Michael,$^{\rm(7)}$
and T.~Kishimoto$^{\rm(8)}$}
\begin{instit}
$^{\rm(1)}$University of Colorado, Department of Physics,
Boulder, Colorado 80309
\end{instit}
\begin{instit}
$^{\rm(2)}$Brookhaven National Laboratory, Upton, New York 11973
\end{instit}
\begin{instit}
$^{\rm(3)}$TRIUMF, 4004 Wesbrook Mall, Vancouver, BC V6T 2A3, Canada
\end{instit}
\begin{instit}
$^{\rm(4)}$Loma Linda University, Loma Linda, California 92350
\end{instit}
\begin{instit}
$^{\rm(5)}$Supercomputer Computations Research Institute, Tallahassee, FL 32306
\end{instit}
\begin{instit}
$^{\rm(6)}$University of Houston, Department of Physics, Houston, TX 77204
\end{instit}
\begin{instit}
$^{\rm(7)}$Ohio University, Department of Physics, Athens, OH 45701
\end{instit}
\begin{instit}
$^{\rm(8)}$Department of Physics, Osaka University, Osaka, Japan 560
\end{instit}
\receipt{28 July 1993}
\begin{abstract}
K$^+$--nucleus quasielastic cross sections measured for a laboratory kaon
beam momentum of 705 MeV/$c$ are presented for 3--momentum transfers of
300 and 500 MeV/$c$.  The measured differential cross sections for C, Ca
and Pb at 500 MeV/$c$ are used to deduce the effective number of nucleons
participating in the scattering, which are compared with estimates based
on the eikonal approximation.  The long mean free path expected for K$^+$
mesons in nuclei is found.  Double differential cross sections for C and Ca
are compared to relativistic nuclear structure calculations.
\end{abstract}

\narrowtext
Quasielastic scattering on complex nuclei probes the nuclear response near the
kinematics for scattering from free nucleons, and is a powerful means
of exploring aspects of nuclear dynamics.
In general, nuclear interactions manifest themselves as
departures from the quasielastic
predictions for a non--interacting Fermi gas of nucleons.
There have been many studies of ($e,e^\prime$) quasielastic scattering, and
interpretations of these have been aided by the well understood nature
and the weakness of the fundamental interaction.
Similar studies with hadronic probes
\cite{hadronica,hadronicb,hadronicc}, including pions \cite{pions},
in principle yield responses not accessible in ($e,e^\prime$)
scattering.  However strong projectile--nucleon interactions
can greatly complicate theoretical analyses of these studies.

In light of these considerations,
the K$^+$ meson is an promising projectile for nuclear quasielastic
scattering studies since it probes responses other than those of the electron
and because the K$^+$--N interaction is
weaker than those of the proton or pion for laboratory momenta below
about 800 MeV/$c$.  Consequently K$^+$ quasielastic
scattering should be more sensitive to the nuclear interior than
is the scattering of any other hadron.
For example, at a laboratory 3--momentum $k$ of 705 MeV/$c$,
the mean free path in nuclear matter is about 4 fm for K$^+$ compared
to about 2 fm for either $\pi^+$ or protons with that momentum.

The weakness of the K$^+$--N interaction has generated much interest
in K$^+$--nucleus elastic scattering, since it was anticipated that simple
multiple scattering treatments of the process should be quite accurate.
However, differential cross section data for K$^+$--C and K$^+$--Ca
elastic scattering
at $k=$ 800 MeV/$c$ are systematically underestimated
by multiple scattering calculations \cite{elastic1,elastic2}.
Similarly, the observed ratio of C to D total cross sections exceeds
predictions \cite{krauss}.  Since multiple scattering calculations
implicitly assume that the bulk of the important absorptive part of the
K$^+$--nucleus scattering potential is due to quasielastic scattering,
direct measurement of this process is likely to be useful in unravelling
the physics of K$^+$--nucleus elastic scattering.

We have measured the quasielastic scattering of K$^+$ mesons from natural
isotopic targets of C, Ca and Pb, and from D in a solid CD$_2$ target.
Double differential cross sections with $k=$ 705 MeV/$c$
were measured at laboratory scattering angles of
24$^\circ$, 33$^\circ$ and 42$^\circ$, corresponding to laboratory
3--momentum transfers $q$ of 300, 400 and 500 MeV/$c$.
At $q=$ 500 MeV/$c$ the kinematic conditions for incoherent quasielastic
scattering are satisfied.
The experiment was done on the C--6 beam line of Brookhaven National
Laboratory using the hypernuclear spectrometer
system (Moby Dick) \cite{mobydick}.  Beam tuning gave nearly equal intensities
of $\pi^+$ and K$^+$ on target, typically around 2 $\times$ 10$^5$
per 1.2 second beam spill each 3.8 seconds.  Trajectories of incident and
scattered particles were determined with position measurements from
15 drift chambers spaced before and after the target.  Particle identification
was carried out with four plastic scintillators, two before
the target and two after, used for time of flight measurements.
In addition, a ${\rm{\check C}}$erenkov counter was used to identify
beam pions.

The momentum acceptance of Moby Dick was measured
with 13 data points from elastic scattering of beam protons from
hydrogen in a solid CH$_2$ target of thickness 1.16 g/cm$^{2}$,
and was checked for consistency with 7 data points from elastic
K$^+$--p scattering from the hydrogen in a nylon target
of thickness 2.89 g/cm$^2$, with known hydrogen composition.
The targets ranged in thickness from 1 to 5 g/cm$^{2}$,
introducing an energy spread in the incident and scattered particles.
In combination with the approximate 3 MeV spectrometer resolution, this
gave an overall energy resolution better than 5 MeV --- quite adequate
for the broad quasielastic peaks observed.  At 24$^\circ$ the resolution
permitted us to make a reliable subtraction of elastic scattering from
the spectra.

{}From three to five spectrometer settings were required to cover
the wide range of excitation energy $\omega$ of each nuclear target.
The finite spectrometer angular acceptance of 3$^\circ$
gave $q$ constant to within approximately 20 MeV/$c$ across the full range
of the spectra.  Measurements of elastic K$^+$--p scattering using the nylon
target, double--checked using another CH$_2$ target of thickness 0.94 g/cm$^2$,
and checked in an additional run using yet another CH$_2$ target
of thickness 2.35 g/cm$^2$,
were normalized to the SP88 phase shift solution \cite{said} to obtain
the cross section scale for the final spectra.  Sample spectra showing the
double differential cross section $d^2\sigma/d\Omega\ d\omega$
versus $\omega$ for C and Ca at $q=$ 500 and 300 MeV/$c$ are
displayed in Figs. \ref{spectrum500} and \ref{spectrum300}.
Only statistical uncertainties are shown in the figures.

To assess the extent to which the K$^+$ probes the nuclear interior
we have extracted ${\rm A}_{\rm eff}$ from our 500 MeV/$c$ data.
${\rm A}_{\rm eff}$ is the
experimentally determined effective number of nucleons ``seen'' by the K$^+$
in quasielastic scattering.  Experimental values have been found by integrating
the double differential cross sections over the quasielastic region.
The resulting $d\sigma /d\Omega$ values appear in Table \ref{glauber}.
The quoted errors in Table \ref{glauber} include both normalization
and systematic uncertainties totalling 12\% and the statistical uncertainty
of 5\%.  ${\rm A}_{\rm eff}$ has been extracted from $d\sigma /d\Omega$
by dividing by the appropriate combination of K$^+$--p and K$^+$--n
differential cross sections.  These cross sections at 42$^\circ$ were
taken to be 2.00 and 1.21 mb/sr, respectively \cite{said}.
The measured H cross section is consistent with the K$^+$--p cross section
and the D cross section agrees with the summed K$^+$--p and K$^+$--n cross
sections.

Table \ref{glauber} also lists ${\rm A}_{\rm eff}^{\rm E}$, the effective
number of nucleons, estimated using the following expression based on the
eikonal approximation:
\begin{equation}
{\rm A}_{\rm eff}^{\rm E} = \int d^2bT(b){\rm e}^{-\sigma_{\rm T}  T(b)}\; ,
\label{differential}
\end{equation}
where
\begin{equation}
T(b) = \int^\infty_{-\infty} dz \rho(\sqrt{b^2+z^2})\; ,
\end{equation}
in which $\rho(r)$ is the nuclear density taken from the ground state
proton matter distributions of Ref. \cite{devries}, and where the neutron
density is assumed to have the same shape as the proton density.
Also $\sigma_{\rm T}$ is the appropriately isospin averaged K$^+$--N
total cross section determined from the K$^+$--p and K$^+$--n total cross
sections of 12.40 and 15.76 mb, respectively \cite{said}.  The experimental
values of ${\rm A}_{\rm eff}$ for C, Ca and Pb are consistently about 30\%
larger than the corresponding eikonal predictions ${\rm A}_{\rm eff}^{\rm E}$.
These results imply, in the context of Eq. (\ref{differential}), that a value
of $\sigma_{\rm T}$ smaller than that given by the phase shift solution
is required by experiment.  Now consider the eikonal expression for the
K$^+$--nucleus total cross section
\begin{equation}
\sigma_{\rm tot} =
2 \int d^2b [ 1 - {\rm e}^{-{1 \over 2}\sigma_{\rm T}  T(b)} ] \; .
\label{total}
\end{equation}
As mentioned above, multiple scattering theory underestimates nuclear
elastic scattering data, including the total cross sections, which
indicates that a larger value of $\sigma_{\rm T}$ is required in
Eq. (\ref{total}).
At present we can offer no explaination of this apparent
contradiction with the quasielastic results.

Independent of their physical origin, the large values of ${\rm A}_{\rm eff}$
found for C, Ca and Pb confirm the nuclear penetration anticipated for K$^+$.
For example we find that the effective number of protons in Pb is 18.1 out
of 82, while for pion single charge exchange (SCX) on Bi the effective
number of protons was found to be only 9.3 out of 83 \cite{ouyang}.
Shown also in Table \ref{glauber} are the radius $R$ and the fraction of
the central density $\rho (R)/\rho(0)$ for which
\begin{equation}
{\rm A}_{\rm eff}/{\rm A} = \int_{R}^\infty d^3r \rho(r) \; .
\end{equation}
Using this simple model we find that K$^+$'s reach a nuclear density
in Pb which is 55\% of the central density, while for pion SCX only
25\% of the central density is reached.

We now compare the data with both nuclear matter and full finite nucleus
calculations based on quantum hadrodynamics (QHD), a relativistic theory of
nuclear dynamics \cite{qhd}.  Further, we present calculations with and without
random phase approximation (RPA) treatment of long range correlations.
Details of our RPA method are found in
Refs.\cite{rpa1,rpa2}.  We have specifically used the relativistic Hartree
approximation (RHA)--RPA in which 1--loop vacuum polarization effects
are included via a local--density--approximation.
Mean field theory (MFT)--RPA calculations, which ignore vacuum polarization
effects, give results quite similar to the RHA--RPA results.
Our RHA calculations employ the meson masses and coupling constants of
Ref. \cite{rhaparam} while the MFT results use the parameters of
Ref. \cite{mftparam}.
In the plane wave impulse approximation, the K$^+$--nucleus double
differential cross section is
\widetext
\begin{equation}
{d^2 \sigma \over d\Omega d\omega} = \kappa \Biggl\{
|f_s|^2\,S_{\rm SS} + |f_v|^2
\Biggl[ \Biggl( {E+E^\prime \over 2m} \Biggr)^2
\Biggl( {Q^4\over q^4} S_{\rm 00} + {Q^2\over q^2} S_{\rm 11} \Biggr)
 - \Biggl( 1 + {Q^2 \over 4m^2} \Biggr) S_{\rm 11}\Biggr]
 + 2 {\rm Re} (f_sf_v^*) {E+E^\prime \over 2m} {Q^2\over q^2} S_{\rm S0}
\Biggr\}\; ,
\end{equation}
\narrowtext\noindent
where $\kappa$ is a kinematical factor arising from the transformation
of the solid angle from the center--of--momentum (c.m.) frame to the
laboratory frame, $Q^2 \equiv q^2-\omega^2$,
$E(E^\prime)$ is the initial (final) K$^+$ total energy, $m$ is the K$^+$
mass and $f_s$ and $f_v$ are relativistic invariants.  These are related
to the on-shell, isospin averaged, c.m. frame K$^+$--N scattering amplitude via
\begin{equation}
f= F + \sigma_n G = {\bar u}_f \bigl[ f_s +
{ \gamma\cdot K \over m}
f_v \bigr] u_i\; ,
\end{equation}
where the $u_i$ ($u_f$) are the initial (final) nucleon Dirac spinors,
$F$ and $G$ are the non--spin--flip and spin--flip scattering amplitudes,
respectively, taken from Ref.\cite{said}, $K^\mu$ is the average
K$^+$ 4--momentum and
$\sigma_n\equiv\vec\sigma\cdot\vec K\times\vec q/|\vec K\times\vec q|$.
Finally, the nuclear responses are
\begin{equation}
S_{ij} = - {1\over \pi}\,{\rm Im}\,\mbox{Tr}\,\bigl[
\theta_i \Pi (\omega,q) \theta_j \bigr] \; ,
\end{equation}
where $\Pi$ is the nuclear polarization insertion.  For $i\rightarrow s$,
$\theta_i \rightarrow \mbox{\bf 1}$, while for $i\rightarrow \mbox{0}$,
$\theta_i \rightarrow \gamma_0$ and for $i \rightarrow \mbox{1}$,
$\theta_i \rightarrow {\vec \gamma}\cdot{\hat e}_T$, where ${\hat e}_T$
is a unit vector in the scattering plane and perpendicular to ${\vec q}$.
All calculations have been scaled by the values of ${\rm A}_{\rm eff}$
listed in Table \ref{glauber}.

While the responses $S_{\rm 00}$ and $S_{\rm 11}$
enter in the Coulomb and transverse ($e,e^\prime$) responses,
the scalar--scalar and mixed scalar--vector responses
$S_{\rm SS}$ and $S_{\rm S0}$ do not appear there.
In addition, due to the detailed nature of the K$^+$--N interaction and
the fact that the responses $S_{\rm SS}$,
$S_{\rm S0}$ and $S_{\rm 00}$ are typically
very similar in magnitude, there are strong cancellations between the sum
of the first two terms and the third term in the expression for the double
differential cross section.  Such cancellations make it possible that
relatively small kinematic or nuclear structure effects might cause strong
departures from naive descriptions of quasielastic scattering.

Figure \ref{spectrum500} presents the $q=$ 500 MeV/$c$ data for Ca and C,
clearly displaying the characteristic shape of a quasielastic response.
Uncorrelated calculations for Ca appear in the upper section of the figure.
The dotted curve is  a relativistic Fermi gas prediction while the
dashed curve includes the effects of a reduced nucleon effective
mass $M^*$ in the nucleus due to the strong attractive scalar
potentials which characterize QHD.  The exact value of the effective mass
used ($M^*/M =$ 0.8576 for Ca in RHA) has been determined self--consistently
at the average nuclear density at which the K$^+$--N interaction occurs.
We note that, while the individual Fermi gas responses each peak near the
$\omega$ given by free K$^+$--N kinematics, the corresponding double
differential cross section peaks at a considerably lower $\omega$ due
to the effects of kinematical factors and cancellations,
as mentioned above.  The considerable influence of such effects on
the shape of the summed response persists in all the calculations discussed
here.
The reduced effective mass used to compute the dashed curve in the upper panel
gives a greatly improved agreement with the data compared to the Fermi gas
result.  In the middle panel RHA--RPA calculations for Ca are displayed.
As shown by the dot--dashed curve, RPA correlations in nuclear matter have
little effect at this momentum transfer.  The solid
curve is a full finite nucleus RHA--RPA calculation
which is remarkably similar to the nuclear matter result, as well as being
in excellent
agreement with data.  The bottom panel shows similar calculations for C
with correspondingly satisfactory accord with experiment.  MFT calculations
are quite similar except that the full finite nucleus results for C are too low
by about 25\%.

Figure \ref{spectrum300} shows the $q=$ 300 MeV/$c$ data for C and Ca.
In this case the quasielastic peak is greatly overshadowed by the strength
concentrated at much lower $\omega$.  The top panel compares uncorrelated
finite nucleus RHA calculations with the C data.  The solid curve in each
panel is the sum
of the isoscalar ($\Delta T= $ 0, dotted) and isovector
($\Delta T= $ 1, dashed) contributions.  Agreement for
$\omega < $ 25 MeV  is poor.  Inclusion of RPA correlations (middle panel)
and the resulting collectivity in the isoscalar response at low $\omega$
corrects this discrepancy to some extent (somewhat stronger collectivity
is seen in the MFT--RPA results).  However for both C and Ca (bottom panel)
collectivity at $\omega < $ 25 MeV is significantly underestimated.
In contrast the data are well described at higher $\omega$.

We conclude that, at the present beam momentum of 705 MeV/$c$, the K$^+$ meson
does probe the nuclear interior more deeply than other hadrons.  In fact,
the effective number of nucleons seen by K$^+$'s  at a momentum transfer of
500 MeV/$c$ is substantially
greater than estimates based on the eikonal
approximation and employing K$^+$--N interactions taken from phase shift
solutions.  This finding is in conflict with studies of K$^+$--nucleus
elastic scattering.  Microscopic QHD calculations of the K$^+$ nuclear
response at $q=$ 500 MeV/$c$ show that $M^*$ effects as well as
RPA correlations and full
treatment of finite nuclear size effects are important in obtaining
a quantitative description of the prominent quasielastic peak.
At $q=$ 300 MeV/$c$ collective strength at low excitation energies
dominates the reponse and is not fully accounted for by the calculations.
At excitation energies above about 25 MeV, full finite nucleus RHA--RPA
calculations agree well with the largely featureless data.

\acknowledgments

This research was supported in part by the
National Science Foundation and in part by the
U.S. Department of Energy under contracts No.~DE-FG02-86ER40269
and DE-AC02-76CH00016.
We would also like to acknowledge the scientists and staff at
Brookhaven National Laboratory and E.~V. Hungerford, W. Jameson, K. Hicks
and M. Holcomb for their assistance.  We are grateful
to the U. S. Japan Cooperative Science Program of the Japanese
Society for the Promotion of Science for their financial support.

\figure{
Quasielastic spectra for Ca and for C
in 4 MeV bins at $q=$ 500 MeV/$c$ are displayed.  Only statistical
uncertainties are shown.  In the top panel the Ca spectrum is compared
to uncorrelated RHA nuclear matter calculations using
$M^*/M = $ 1 (dotted) and $M^*/M = $ 0.8576 (dashed).
In the middle panel the Ca spectrum is compared to nuclear matter
(dot--dashed) and to finite nucleus (solid) RHA--RPA calculations.
A similar comparison with the C spectrum appears in the bottom panel.
Arrows indicate the location of free K$^+$--nucleon scattering.
\label{spectrum500}}

\figure{
Quasielastic spectra for C and for Ca
in 3 MeV bins at 300 MeV/$c$ are displayed.  Only statistical
uncertainties are shown and elastic peaks have been removed.
Finite nucleus RHA calculations for C appear in the top panel.
Finite nucleus RHA--RPA calculations for C are shown in the middle panel.
The solid lines are the sums of the isoscalar (dotted) and isovector
(dashed) responses in each panel.
Finite nucleus RHA--RPA calculations for Ca are presented in the bottom panel.
Arrows indicate the location of free K$^+$--nucleon scattering.
\label{spectrum300}}

\newpage
\widetext
\begin{table}
\caption{The measured differential cross sections for K$^+$--nucleus
quasielastic scattering at 500 MeV/$c$ are listed.
${\rm A}_{\rm eff}$ is the experimentally determined effective
number of nucleons participating in the scattering, and is compared to
${\rm A}_{\rm eff}^{\rm E}$, deduced from the
eikonal approximation for C, Ca and Pb.  Also the radius
$R$ and fraction of the central nuclear density reached by the K$^+$
are shown for C, Ca and Pb.
}
\label{glauber}
\begin{tabular}{lcccccc}
  & ${d\sigma / d\Omega}$ (mb/sr) &
${\rm A}_{\rm eff}$ & ${\rm A}_{\rm eff}^{\rm E}$ &
${\rm A}_{\rm eff}/{\rm A}_{\rm eff}^{\rm E}$ &
 $R$(fm) & $ \rho(R) / \rho(0) $ \\
\tableline
H  & \dec  2.10 $\pm$\dec 0.27 & \dec  1.05 $\pm$\dec 0.14 &
          &  &  &  \\
D  & \dec  3.38 $\pm$\dec 0.44 & \dec  2.10 $\pm$\dec 0.27 &
          &  &  &  \\
C  & \dec 13.6 $\pm$\dec  1.8  & \dec  8.47 $\pm$\dec 1.1  &
\dec  6.7 & \dec 1.27 $\pm$ 0.17 & \dec 1.7 & \dec 0.90 \\
Ca & \dec 33.7 $\pm$\dec  4.4  & \dec 21.0 $\pm$\dec 2.7  &
\dec 16.0 & \dec 1.31 $\pm$ 0.17 & \dec 3.2 & \dec 0.64 \\
Pb & \dec 69.5 $\pm$\dec  9.0  & \dec 45.7 $\pm$\dec 5.9  &
\dec 36.0 & \dec 1.27 $\pm$ 0.17 & \dec 6.5 & \dec 0.55 \\
\end{tabular}
\end{table}
\narrowtext
\end{document}